\documentclass[prl,aps,twocolumn,showpacs,showkeys]{revtex4} 
\usepackage{graphicx} 

\begin{document} 
\newcommand{\p}{\partial} 
\newcommand{\ls}{\left(} 
\newcommand{\rs}{\right)} 
\newcommand{\beq}{\begin{equation}} 
\newcommand{\eeq}{\end{equation}} 
\newcommand{\beqa}{\begin{eqnarray}} 
\newcommand{\eeqa}{\end{eqnarray}} 
\newcommand{\bdm}{\begin{displaymath}} 
\newcommand{\edm}{\end{displaymath}} 
 
\title{Isospin effects on sub-threshold kaon production at intermediate 
energies} 
\author{G. Ferini$^1$, T. Gaitanos$^2$, M. Colonna$^1$, M. Di Toro$^1$,
  H.H. Wolter$^2$} 
\affiliation{
$^1$ Universit\`a di Catania and INFN, Lab. Nazionali
del Sud, 95123 Catania, Italy\\
$^2$ Dept. f\"ur Physik, Universit\"at M\"unchen, D-85748 Garching, Germany\\
E-mail: ditoro@lns.infn.it}

\begin{abstract} 
We show that in collisions with neutron rich heavy ions at energies 
around the production threshold $K^0$ and $K^+$ yields might 
probe the isospin dependent part of the nuclear
Equation of State ($EoS$) at high baryon densities. In particular
 we suggest the
$K^0/K^+$ ratio as a promising observable. Results obtained in a fully 
covariant relativistic transport approach are presented for central $Au+Au$ 
collisions in the beam energy range $0.8-1.8~AGeV$. The focus is put 
on the $EoS$ influence which goes beyond the $collision-cascade$ picture. 
The isovector part
of the in-medium interaction affects the kaon multiplicities
via two mechanisms: i) a {\it symmetry potential} effect, i.e. a 
larger neutron repulsion
in n-rich systems (isospin fractionation); ii) a {\it threshold} effect,
due to the change in the self-energies of the particles involved in inelastic
processes. Genuine relativistic contributions are revealed, that could allow
to directly ``measure'' the Lorentz structure of the effective isovector
interaction. 

\end{abstract}
\maketitle 
 
For at least two decades particle production has been suggested as a  
useful tool to constrain the poorly known high density behaviour of the  
nuclear equation of state ($EoS$) \cite{StockPR135,AichkoPRL55}. 
This information is crucial both 
to learn about the effective nuclear interaction and to get a deeper 
insight into many interesting astrophysical phenomena. 
In particular pion and (subthreshold) kaon productions have been extensively 
investigated both theoretically 
\cite{AichkoPRL55,HartNPA580,MaruNPA573,KoLiNPA575,KoLiPLB349,FuchsPRC56,FuchsPRL86} and  
experimentally \cite{Kaos94,Kaos9901,Kaos05}, leading to the conclusion of 
a soft 
behaviour of the $EoS$ 
at high densities, see the recent refs. \cite{FuchsPPNP56,HartPRL96}. 
Kaons ($K^0,K^+$)
appear as particularly sensitive probes since they are produced in the high 
density phase almost without subsequent reabsorption effects. At variance, 
antikaons ($\bar K^0,\bar K^-$) are strongly coupled to the hadronic medium 
through strangeness
exchange reactions \cite{FuchsPPNP56,CassingNPA727,WeberPRC67}.
In this paper we show that the isospin dependence of the $K^{0,+}$
production cross sections can be also used to probe the isovector part of
the $EoS$: we propose the $K^0/K^+$ yield ratio as a good observable to
constrain the high density behavior of the symmetry energy,
 $E_{sym}$, \cite{Isospin01,BaranPR410}. 

In heavy ion collisions ($HIC$s) at intermediate energies (1-2 AGeV) a 
transient state of 
highly compressed and heated matter is created, with density up to 3-4 times  
the saturation value.   
Several dynamical observables have hence been proposed to  
characterize the density dependence of  $E_{sym}$, such as     
charge ratios and
collective isospin flows for highly  
energetic nucleons and pions 
\cite{BaoNPA708,GrecoPLB562,GaitNPA732,BaoPRC71,QLiPRC72,QLiJPG32} and isospin 
transparency as expressed in  
transport ratios \cite{GaitPLB595,ChenPRC69,QLinth0603}. Controversial
deductions arise from different relativistic  and non-relativistic models, 
see \cite{BaranPR410}. 

We show that,
within a covariant description of nuclear dynamics,
 from the $K^{0}/K^{+}$ ratios we can directly investigate
the Lorentz structure, i.e. the scalar-vector decomposition, of the isovector 
sector of the effective in-medium hadron Lagrangian. 
Some promising indications have been
recently obtained in nuclear matter calculations \cite{FeriniNPA762}, here we
present results for realistic {\it open} systems, i.e. for collisions
of neutron-rich heavy ions in the energy range around the kaon production
threshold ($1.56~AGeV$).

Using a transport model, derived within the Relativistic Mean Field  
approximation ($RMF$) of Quantum-Hadro-Dynamics \cite{QHD}, we analyze 
pion and kaon production in central $^{197}Au+^{197}Au$ collisions in 
the $0.8-1.8~AGeV$
 beam 
energy range, with different $RMF$ effective field choices for 
$E_{sym}$. We will compare results of three Lagrangians with constant 
nucleon-meson 
couplings ($NL...$ type, see \cite{BaranPR410,GaitNPA732}) and one with density
dependent couplings ($DDF$, see \cite{GaitNPA732}), recently suggested 
for apparent better nucleonic properties of neutron stars \cite{Klahn06}.
In order to isolate the sensitivity to the isovector components we use
models showing the same "soft" $EoS$ for symmetric matter 
\cite{BaranPR410,GaitNPA732}.
This is achieved via Non-Linear ($NL...$) contributions of the isoscalar
 scalar $\sigma$ and vector $\omega$ mesons in the constant coupling cases, and
via suitable density dependences in the $DDF$ model.

For the isovector part, in the simple $NL$ choice the mean field is not 
isospin dependent
and the symmetry energy is just due to kinetic (Fermi) contributions.  
The $NL\rho$ model 
contains
an  isovector-vector effective field  
($\rho$ meson), which leads to a splitting of the vector 
self-energies between  
protons and neutrons, and to  a more repulsive force    
experienced by neutrons with respect to protons  in neutron-rich matter  
\cite{BaranPR410}. In the $NL\rho\delta$ model also an   
isovector-scalar $\delta$ field is included, which gives an    
effective mass splitting between protons and neutrons 
\cite{KubisPLB399,LiuPRC65}.


With both $\rho$- and $\delta$-meson couplings, $f_{\rho,\delta}$,
 $E_{sym}$ can be written as \cite{LiuPRC65}

\begin{equation} 
E_{sym} = \frac{1}{6} \frac{k_{F}^{2}}{E_{F}} +  
\frac{1}{2} 
\left[ f_{\rho} - f_{\delta}\left( \frac{m^{*}}{E_{F}} \right)^{2} 
\right] \rho_{B} 
\label{esym3} 
\quad . 
\end{equation} 

leading to a partial cancellation of scalar and vector components, equivalent
to that of the $\sigma$ and $\omega$ fields in the isoscalar sector.
Since the $\delta$ field couples to the scalar density, $f_\delta$ is
multiplied with a density dependent quenching factor in Eq.(\ref{esym3}).
It is then clear that the increase by nearly a factor of three in $f_{\rho}$
in the $NL\rho\delta$ model is necessary to compensate the attraction 
due to the 
scalar $\delta$ field, in order to get the same bulk asymmetry parameter   
$a_4=30.5MeV$ at saturation, \cite{LiuPRC65,GaitNPA732}. 
Consequently the inclusion of a $\delta$ field leads to a 
stiffer symmetry energy at high baryon densities \cite{LiuPRC65}, and to 
larger vector self-energies for nucleons, as discussed in 
\cite{GrecoPLB562,FeriniNPA762}.
In the $DDF$ model
the $f_{\rho}$ is exponentially decreasing with density, resulting in a 
rather "soft" 
symmetry term at high density \cite{GaitNPA732,Klahn06}. 

We like to note the $genuine$ relativistic 
nature of these mechanisms that will directly influence transport dynamics
and particle production. A transport code with only relativistic 
kinematics combined to density dependent interactions will miss
these effects. 

 In the energy range considered here the 
nucleon-nucleon inelastic 
channels can be restricted to the 
excitation of the lowest mass resonance $\Delta(1232)$ and perturbative  
kaon ($K^{+,0}$) production through baryon-baryon collisions
 $BB \longrightarrow BYK$, where $B$ stands  
for nucleons or resonances and $Y$ for hyperons  
($\Lambda,\Sigma^{\pm,0}$).
Pions are produced via the decay of the $\Delta(1232)$ resonance and - after
propagation and rescattering - can contribute 
to the kaon yield through 
collisions with baryons: 
$\pi B \longrightarrow YK$. All these processes are treated within
 a relativistic 
hadronic transport model of Boltzmann-Uehling-Uhlenbeck type 
($RBUU$), i.e. including a hadron mean field propagation 
\cite{BlattelRPP56,FuchsNPA589}.
The latter point, which goes beyond the ``collision cascade'' picture, is
essential for particle production yields since it directly affects the
energy balance of the inelastic channels, as extensively analysed in 
the following. 
Further details can be found in ref.
\cite{FeriniNPA762}, where all
the parametrizations used for the corresponding cross sections are also
specified. 

Due to the $K^{+,0}$ long mean free path their in-medium widths
can be expected to be small and an on-shell quasiparticle treatment
appears suitable. Here, as in \cite{FeriniNPA762}, we do not generally
use potentials for kaons and propagate them as free particles. 
This simplification is justifiable, since our 
main aim here is to show how 
the Lorentz structure of the symmetry energy influences the $K^{0}/K^{+}$ 
ratio, which should not sensitively depend on the kaon-nucleon potential
as this appears roughly the same (slightly repulsive) for both 
 $K^{0}$ and $K^{+}$ \cite{FuchsPPNP56}. 

\begin{figure}[t] 
\includegraphics[scale=0.35]{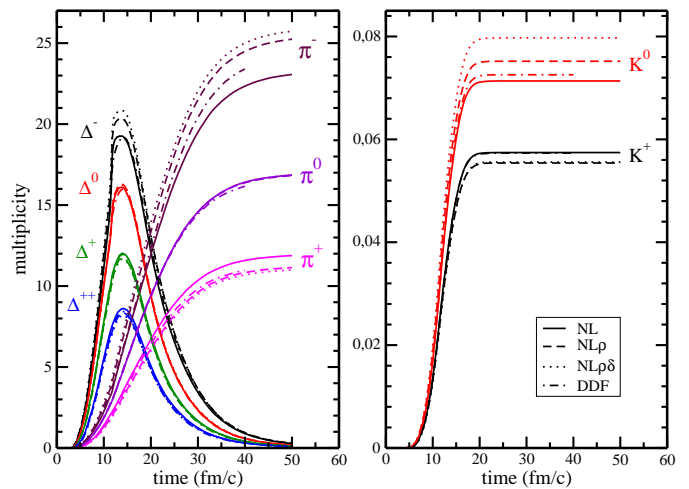} 
\caption{Time evolution of the $\Delta^{\pm,0,++}$ resonances (left) and  
the pions $\pi^{\pm,0}$ (right) for a central ($b=0$ fm impact parameter)  
Au+Au collision at 1 AGeV incident energy. Transport calculation using the  
$NL, NL\rho, NL\rho\delta$ and $DDF$ models for the iso-vector part of the  
nuclear EoS are shown.  
} 
\label{fig1} 
\end{figure}

Fig. \ref{fig1} reports  the temporal evolution of $\Delta^{\pm,0,++}$  
resonances and pions ($\pi^{\pm,0}$) (left panel) and kaons ($K^{+,0}$) 
(right panel) 
for central Au+Au collisions at 1 AGeV. Results are shown for all the 
tested models.

It is clear that, while the pion yield freezes out at times of the order of 
$50 fm/c$, i.e. at the final stage of the reaction (and at low densities),
kaon production occurs within the very early stage of the reaction,
and the yield saturates at around $20 fm/c$. Kaons are then suitable 
to probe the 
high density phase of nuclear matter. 
This is  not the case for
pions, which suffer from reabsorption and isospin exchange processes that 
modify both the absolute primordial yield and the $\pi^{-}/\pi^{+}$  
ratio. 
In fact from Fig. \ref{fig1} we see that the results for different
models are rather similar, and thus
pion multiplicities depend only weakly on the  
isospin part of the nuclear mean field.
However, a slight increase (decrease) in the $\pi^{-}$ ($\pi^{+}$) 
multiplicity is observed when going from the $NL$ to the $NL\rho$ and then to
the $NL\rho\delta$ model, i.e. increasing the vector contribution $f_\rho$
in the isovector channel. This trend is 
more pronounced for kaons, see Fig. \ref{fig1}
right panel, due to the high density selection of the source and the
proximity to the production threshold. The results for the $DDF$ model,
density dependent couplings with a large $f_{\rho}$ decrease at high density,
are fully consistent. They are always closer to the $NL$ case (without
isovector interactions) but the difference still seen for $\pi^{+,-}$
is completely disappearing for $K^{0,+}$, selectively produced at high 
densities. 

When isovector fields are included the symmetry potential energy in 
neutron-rich matter is repulsive for neutrons and attractive for protons.
In a $HIC$ this leads to a fast, pre-equilibrium, emission of neutrons
towards more symmetric nuclear matter. Such a $mean~field$ mechanism, which 
has been
extensively investigated, often referred to as isospin fractionation
\cite{Isospin01,BaranPR410}, is responsible for a reduction of the neutron
to proton ratio during the high density phase, with direct consequences
on particle production in inelastic $NN$ collisions.

Threshold effects represent a more subtle question. The energy conservation in
a hadron collision in general has to be formulated in terms of the canonical
momenta, i.e. for a reaction $1+2 \rightarrow 3+4$ as
\begin{equation}
s_{in} = (k_1^\mu + k_2^\mu)^2 = (k_3^\mu + k_4^\mu)^2 = s_{out}
\label{encan}
\end{equation}
Since hadrons are propagating with effective (kinetic) momenta and masses,
 an equivalent equation should be formulated starting from the effective
in-medium quantities $k^{*\mu}=k^\mu-\Sigma^\mu$ and $m^*=m+\Sigma_s$, where
$\Sigma_s$ and $\Sigma^\mu$ are the scalar and vector self-energies, 
respectively, depending on the isovector channel structure.
In particular, for the general $\sigma\omega\rho\delta$ case one obtains for 
the self-energies 
of protons and neutrons:
\begin{eqnarray}
\Sigma_{s}(p,n) & = & - f_{\sigma}\rho_{s} \pm f_{\delta}\rho_{s3} 
\label{sigs}\\
\Sigma^{\mu}(p,n) & = & f_{\omega}j^{\mu} \mp f_{\rho}j^{\mu}_{3},
\label{sigv}
\quad .
\end{eqnarray}
(upper signs for neutrons), 
where $\rho_{s}=\rho_{sp}+\rho_{sn},~
j^{\alpha}=j^{\alpha}_{p}+j^{\alpha}_{n},\rho_{s3}=\rho_{sp}-\rho_{sn},
~j^{\alpha}_{3}=j^{\alpha}_{p}-j^{\alpha}_{n}$ are the total and 
isospin scalar 
densities and currents and $f_{\sigma,\omega,\rho,\delta}$  are the coupling 
constants of the various 
mesonic fields, \cite{nonlinear}.

In reactions where nucleon resonances, especially the different isospin
states of the $\Delta$ resonance, and hyperons enter, also their self
energies are relevant for energy conservation. We specify them in the usual
way according to the light quark content and with appropriate Clebsch-Gordon
coefficients \cite{FeriniNPA762}. 

In the most general case the isovector scalar and vector self-energies
enter the new threshold condition for a given 
inelastic process \cite{FeriniNPA762}
\begin{equation} 
s_{in}\geq {(m_{3}^{*}+\Sigma _{3}^{0}+m_{4}^{*}+
\Sigma _{4}^{0})^2 - 
({\bf \Sigma}_{3}+{\bf \Sigma}_{4})^2}
\label{thresh}
\quad .
\end{equation}

The condition of energy conservation in inelastic hadron collisions 
will
influence the particle production in two different ways. On one hand it will 
directly determine the thresholds and thus the multiplicities of a certain type
of particles, in particular of the sub-threshold ones, as 
here for the kaons. Secondly it may favour or penalize reactions, because
the self-energies in the final channel are more attractive or repulsive than
in the initial one, and consequently the phase space in the final channel
is larger or smaller.

Hence, we clearly see that the competition between scalar 
and vector
isovector fields is responsible for the isospin fractionation $and$
 modifies particle production rates. In 
particular while the scalar and vector isovector fields tend to cancel in the
symmetry term, see
 Eq.(\ref{esym3}), they can have very different dynamical effects, as 
already noted in the flow analysis of ref. \cite{GrecoPLB562}. 
In fact {\it mean field} 
and {\it threshold} effects
are acting in opposite directions on particle production  and might 
compensate each other.

 As an example, $nn$
collisions excite $\Delta^{-,0}$ resonances which decay mainly to $\pi^-$.
 In a
neutron-rich matter the mean field effect pushes out neutrons making the 
matter more symmetric and thus decreasing the $\pi^-$ yield. The threshold 
effect on the other hand is increasing the rate of $\pi^-$'s due to the
enhanced production of the $\Delta^-$ resonances: 
now the $nn \rightarrow p\Delta^-$ process is favored
(with respect to $pp \rightarrow n\Delta^{++}$) 
 since more effectively a neutron is converted into a proton.

We have to note that such interplay between the two mechanisms cannot be 
fully included in a non-relativistic dynamics,
in particular in calculations where the baryon symmetry potential is
treated classically \cite{BaoNPA708,BaoPRC71,QLiPRC72,QLiJPG32}.
A typical example is the strength of the isovector-vector $\rho$-coupling
which is linked to the symmetry energy but it is largely varying with the
Lorentz structure of the isovector interaction. 
Actually aim of this paper is to show
that the data could be used to directly probe the
covariant structure of the effective interaction in the isovector channel
at high baryon densities.

In the left panel of Fig. \ref{fig1} we see that $\Delta^-$ (and $\pi^-$)
production increases  when going in the direction of
larger $f_\rho$ couplings 
$NL \rightarrow DDF \rightarrow NL\rho \rightarrow NL\rho\delta$. 
As discussed before,
 the isospin fractionation
lowers the $n/p$ ratio and thus decreases the $\Delta^-$ (and $\pi^-$)
yield, while the threshold mechanism increases it due to the larger 
$nn$ self-energies. In fact the latter effect appears to overcompensate 
the former.
The opposite is true for the $\Delta^{+,++}$ (and $\pi^+$) production.
 
As seen in the right panel of Fig. \ref{fig1}, a similar argument
holds for $K^{0}$ and $K^{+}$ mesons, which mainly come from
$nn$ (or $\pi^{-}n$) and $pp$ (or $\pi^{+}p$) collisions, respectively, and 
thus exhibit the same trend 
as $\pi^{-}$ and $\pi^{+}$. However, the isospin effect in this 
case is more 
pronounced because the changes in the self energies for the different
 models play a more crucial role close to
the kaon production threshold. 

\begin{figure}[t] 
\includegraphics[scale=0.38]{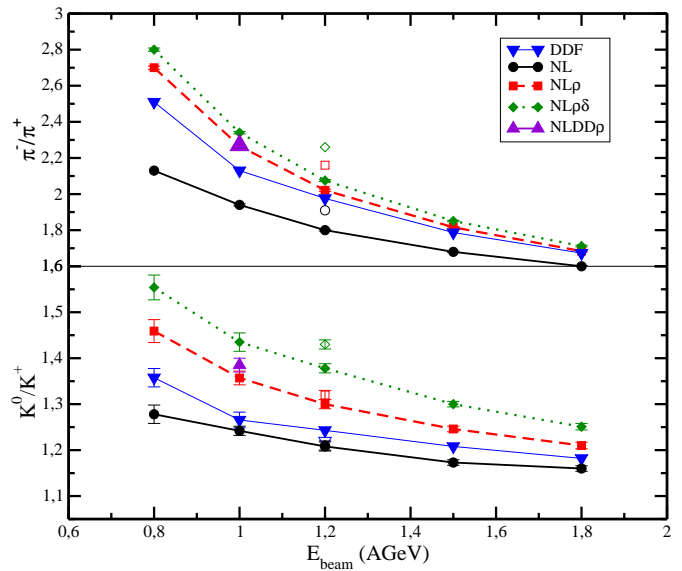} 
\caption{
$\pi^-/\pi^+$ (upper) and  $K^{+}/K^{0}$ (lower) ratios
as a function of the incident energy for the same reaction and models
as in Fig. \ref{fig1}.
In addition we present, for $E_{beam}=1~AGeV$, $NL\rho$ results with a density 
dependent
$\rho$-coupling (triangles), see text. The $open$ symbols
at $1.2~AGeV$ show the corresponding results for a $^{132}Sn+^{124}Sn$
collision, more neutron rich.
Note the different scale
for the $\pi^-/\pi^+$ ratios.
} 
\label{fig2} 
\end{figure}

Finally the beam energy dependence of the $\pi^-/\pi^+$ (left) and 
$K^{0}/K^{+}$ 
(right) ratios is shown in Fig. \ref{fig2}. At each energy we see an
increase of the yield ratios with the models
$NL \rightarrow DDF \rightarrow NL\rho \rightarrow NL\rho\delta$. 
The effect is larger for the $K^{0}/K^{+}$ compared to the $\pi^-/\pi^+$
ratio. This is due to the subthreshold production and to the fact that
the isospin effect enters twice in the two-steps production of kaons, see
\cite{twosteps}. Between the two extreme $DDF$ and
$NL\rho\delta$ isovector interaction models, the 
variations in the ratios are of the order of $14-16 \%$ for kaons, while 
they reduce to about $8-10 \%$ for pions.
Interestingly the Iso-$EoS$ effect for pions is increasing at lower energies,
when approaching the production threshold.

We have to note that in a previous study of kaon production in excited nuclear
matter the dependence of the $K^{0}/K^{+}$ yield ratio on the effective
isovector interaction appears much larger, about ten times more for a
system with the $Au$ asymmetry (see Fig.8 of ref.\cite{FeriniNPA762}).
The point is that in the non-equilibrium case of a heavy ion collision
the asymmetry of the source where kaons are produced is in fact reduced
by the $n \rightarrow p$ ``transformation'', due to the favored 
$nn \rightarrow p\Delta^-$ processes. This effect is almost absent at 
equilibrium due to the inverse transitions, see Fig.(3) of 
ref.\cite{FeriniNPA762}. Moreover in infinite nuclear matter even the fast
neutron emission is not present. 

 Thus in an $open$ system the signal appears weaker but still accessible 
 experimentally, if it can be shown to be
robust. 
In order to further stress the distinction between effects of
 the stiffness of the symmetry energy and the detailed Lorentz structure
of the isovector part of the effective Lagrangian, we also show
the results for the $K^{0}/K^{+}$ with another parametrization of
$E_{sym}$. This model, $NLDD\rho$, is
a variant of $NL\rho$
with a density dependent $\rho$-coupling, built
in such a way as to reproduce the same stiffer
$E_{sym}(\rho_B)$ of the $NL\rho\delta$ model 
(see also ref. \cite{GrecoPLB562}).
 The results for the  
$\pi^-/\pi^+$ and  $K^{0}/K^{+}$ ratios are shown in Fig. \ref{fig2} for
$E_{beam}=1.0~AGeV$ as triangles. We see that they are closer to the 
$NL\rho$ results (with a constant $f_{\rho}$) than to 
the ones of the $NL\rho\delta$ choice which has the
same iso-stiffness. This nicely confirms that the differences observed 
going from the $NL\rho$ to the  $NL\rho\delta$ parametrization are not due to
 the slightly increased stiffness of $E_{sym}(\rho_B)$, but more specifically
to the competition between the attractive scalar $\delta$-field and the
repulsive vector $\rho$-field in the isovector channel, which leads to the
increase of the vector coupling (see the comments
to the Eq.(\ref{esym3})). Hence the differences between the vector 
self-energies
of neutrons and $\Delta^-$ resonances vs. the proton and $\Delta^{++}$ ones
are greater in the $NL\rho\delta$ model than in other models without scalar
isovector fields and the production cross sections are consequently affected.
As a result, the subthreshold $K^0~(K^+)$ yield is sensitively enhanced
 (reduced) in the $NL\rho\delta$ case.

 In the same Fig. \ref{fig2} we also report results at $1.2~AGeV$ for
the $^{132}Sn+^{124}Sn$ reaction, induced by a radioactive beam, with
an overall larger asymmetry (open symbols). The isospin effects are 
clearly enhanced.
Finally we have performed some calculations including kaon potentials,
derived from a chiral lagrangian as suggested 
in \cite{FuchsPPNP56}.
The $K^{0}/K^{+}$ ratio appears not very sensitive to this in-medium 
modification of the
kaon single particle energies. This is somewhat expected since we have
a similar small repulsive contribution for both $K^{0,+}$ mesons.

From the present discussion we conclude that subthreshold kaon production
could provide a promising tool to extract information on the isovector part
of the nuclear interaction at high baryon density. We have seen that, at 
beam energies 
below and around
the kinematical threshold, the $K^{0}/K^{+}$ inclusive yield ratio is  more 
sensitive to the Lorentz structure of $E_{sym}$ than the $\pi^{-}/\pi^{+}$.

 We would like to stress the two most important results of 
our study: 
i) We have shown that
isospin effects are important not only at the mean-field
level (isospin fractionation) but they also influence significantly 
particle production cross sections. As a matter of fact, we observe
that, for the reactions studied here, the modifications induced in
 the inelastic vertices represent the dominant effects.   
ii)  At relativistic energies, due to the Lorentz structure of the
isovector nuclear interaction, the isotopic content of particle
emission is not directly related to the symmetry energy value, but it
can be rather considered as a measure of the strength of isovector 
vector channel. 
 
We note that the isospin effects on the kaon inclusive yield ratios
at the freeze-out appear not too strong, although experimentally accessible. 
It seems
important to select more exclusive kaon observables, in particular with 
a trigger related to 
an early time $K$-production.  A transverse momentum selection of
pion yields, corresponding to a higher density source, should also be 
rather sensitive to isospin effects, in particular
at lower energies, closer to the production threshold.
A large asymmetry of the colliding matter
is in any case of relevance. 
In this sense our
work strongly supports the study of particle production at the
new relativistic radioactive beam facilities.

\section*{Acknowledgments} 
We would like to thank Christian Fuchs for providing us with the 
kaon package  
subroutines of the T\"{u}bingen group and for valuable discussions.  
 
 


\begin{thebibliography}{99}
\bibitem{nonlinear}
For the $NL$ models in Eq.(\ref{sigs}) the scalar density  should be 
replaced by
$\sigma(\rho_{s})$, solution of the non linear 
equation for the $\sigma$ field \cite{LiuPRC65,BaranPR410}.
\bibitem{twosteps}
In the energy 
range explored here, the main contribution to the kaon yield comes from the
pionic channels, in particular from $\pi N$ collisions, and from the
 $N \Delta$ channel, which together account 
for nearly $80 \%$ of the total yield, see \cite{FeriniNPA762}.
\end{thebibliography}
\end{document}